\documentclass[preprint,prd,nofootinbib,preprintnumbers]{revtex4}
\usepackage{graphicx, bm}

\begin{document}
\begin{flushright}
OSU-HEP-15-02\\ UMD-PP-015-005
\end{flushright}
\vspace*{0.2in}

\begin{center}
{\Large\bf Limiting Lorentz Violation from \\ Neutron--Antineutron Oscillation}\end{center}

\begin{center}{
 {\bf K.S. Babu$^1$ and Rabindra N. Mohapatra$^2$}}

 {\it $^1$Department of Physics, Oklahoma State University, Stillwater, OK, 74078, USA}

{\it $^2$Maryland Center for Fundamental Physics and Department of Physics,
University of Maryland, College Park, Maryland 20742, USA}
\end{center}
\date{\today}

\vspace*{0.2in}

\begin{abstract}
We point out that if neutron--antineutron oscillation is observed in a free neutron oscillation experiment, it will put an upper limit on the strengths of  Lorentz invariance violating (LIV) mass operators for neutrons at the level of $10^{-23}$ GeV or so, which would be the most stringent limit for neutrons. We also study constraints on $\Delta B=2$ LIV operators and find that for one particular operator degaussing is not necessary to obtain a visible signal. We also note that observation of $n-\bar{n}$ oscillation signal in the nucleon decay search experiment involving nuclei does not lead to any limit on LIV operators since the nuclear potential difference between neutron and antineutrons will mask any Lorentz violating effect.
\end{abstract}

\maketitle

\section{Introduction}
Possible violation of Lorentz invariance has been a topic of  much  theoretical interest for the past two decades~\cite{alan,other} and has been followed up by many experimental searches in the laboratory as well as in the domain of astrophysics. This violation can occur for photons, neutrinos, atoms as well as hadrons such as kaons~\cite{kkbar}, protons and neutrons~\cite{neutrons}. The current limits from various processes are summarized in ~\cite{alan2}. In this paper, we discuss constraints that can be derived, if neutron to antineutron oscillation is observed, since presence of certain kinds of baryon number conserving and Lorentz invariance violating (LIV) operators in the Lagrangian lead to suppression of the oscillation (see below for details).

Neutron--antineutron  ($n-\bar{n}$) oscillation~\cite{nnbar} is a phenomenon where in vacuum and in the limit of small magnetic fields,  a neutron spontaneously converts to an antineutron. It provides an alternative mode of baryon number violation compared to the well known and well studied proton decay. The main interest in investigation of the  $n-\bar{n}$ transition has been due to two factors: (i) it has a different selection rule for baryon number non-conservation compared to canonical proton decay mode $p\to e^+\pi^0$ and (ii) it probes physics at a much lower scale than proton decay. Furthermore, in contrast to conventional $B-L$ conserving proton decay modes, $n-\bar{n}$ oscillation is more intimately connected to the discussion of origin of matter in the universe~\cite{basy}. An experimental search for $n-\bar{n}$ oscillation with free neutrons was conducted at ILL~\cite{ILL} and an upper limit has been established on the strength for the strength of this transition. Searches for this process in nucleon decay searches have also been carried out~\cite{abe}. Currently, there are plans to search for this process at a higher sensitivity level~\cite{new} using the spallation neutrons at ESS laboratory in Lund, Sweden. It is therefore of interest to investigate whether one can learn anything else about the physics of neutrons from this experiment. In this brief note, we point out that the search for this process can also yield useful information on possible Lorentz invariance violation  for neutrons both in the $\Delta B=0$ and $\Delta B\neq 0$ channels.

This paper is organized as follows. In Sec. II, we take examples of two simple $\Delta B = 0$ LIV operators involving neutrons and discuss how one can obtain a limit on the strength of this operator once $n-\bar{n}$ oscillation is observed. In Sec. III, we introduce examples of baryon number violating neutron operators and study their effect on $n-\bar{n}$ oscillation and show that for certain kind of operators, the $n-\bar{n}$ transition between different spins is actually independent of external magnetic fields. Sec. IV deals with a generalized class of $\Delta B = 0$ LIV operators
and Sec. V with $\Delta B = 2$ LIV operators.  We make some comments in sec. VI and then conclude in sec. VII.

\section{ $\Delta B=0$ LIV operators for neutrons and $n-\bar{n}$ oscillation}

We will use an effective operator formalism in terms of the neutrons rather than the elementary quarks since this is a low energy process. In principle, such LIV operators involving neutrons should originate from LIV operators added to the conventional QCD Lagrangian. At the moment we do not know how to translate precisely the constraints on the strengths of the effective neutron operators to those in the QCD Lagrangian.

We start with the conventional effective Lagrangian for free neutron oscillation to antineutrons:
\begin{eqnarray}
{\cal L}~=~i\bar{n}\gamma^{\mu}\partial_\mu n -m\bar{n}n+ \frac{1}{2}\delta_{B=2} n^TCn+h. c.
\end{eqnarray}
The ILL  $n-\bar{n}$ oscillations search with free neutrons has established  a lower limit on the transition time, $\tau_{n-\bar{n}} \geq 0.8 \times 10^7$ sec \cite{ILL}.  This translates into a limit $\delta_{B=2} \leq 10^{-28}$ GeV. Similar limits on $\delta_{B=2}$ are also obtained from nucleon decay searches ~\cite{abe}.

In order to study the effect of Lorentz invariance violation on $n-\bar{n}$ oscillation, we proceed in steps. First we include two simple
baryon number conserving LIV terms and in a subsequent section, add $\Delta B=2$ terms in the Lagrangian.  The lowest dimensional operators
have either positive mass dimensions or are dimensionless.  We focus only on such operators.  Consider the simple set of $B$-conserving LIV
operators first.
\begin{eqnarray}
{\cal L}_{LIV}~=~ a_\mu \bar{n}\gamma^\mu n -ic_{\mu\nu}\bar{n}\gamma^\mu \partial^\nu n~+~h.c.
\end{eqnarray}
where $a_\mu$ and $c_{\mu\nu}$ are spurion fields. Once we give vacuum expectation values to these fields such that they maintain rotational invariance i.e. $\langle a^0 \rangle =\delta^{(1)}_{LV}$ and
$\langle c^{00} \rangle =\delta^{(2)}_{LV}$, we get
\begin{eqnarray}
{\cal L}_{LIV}~=~ \delta^{(1)}_{LV} n^\dagger n -i\delta^{(2)}_{LV}n^\dagger \partial_0 n~+~h.c.
\end{eqnarray}

First thing to note is that the first term violates both Lorentz invariance as well as CPT whereas the second term $\delta^{(2)}_{LV}$ violates only Lorentz invariance but conserves CPT~\cite{km}. To see the effect of these terms, let us initially focus on the first term. We expand the free field in terms of the creation and annihilation operators for neutron and antineutron as follows, using the formalism to treat spin effects in this case developed in a recent paper~\cite{gardner}:
\begin{eqnarray}
\psi(x)=\sum_{p,s}\sqrt{\frac{m}{EV}}\left[b_{p,s}u(p,s)e^{-ip.x}+d^\dagger_{p,s}v(p,s)e^{ip.x}\right]\\
{\rm where} ~~u(p,s)={\cal N}\left(\begin{array}{c}\chi^{(s})\\\frac{\vec{\sigma}\cdot\vec{p}}{E+m}\chi^{(s)}\end{array}\right);~~v(p,s)={\cal N}\left(\begin{array}{c}\frac{\vec{\sigma}\cdot\vec{p}}{E+m}\chi^{\prime(s)}\\ \chi^{\prime(s)}\end{array}\right);
\end{eqnarray}
with $s = +$ or $-$ along with $\chi^+=\left(\begin{array}{c}1\\0\end{array}\right)$  and $\chi^-=\left(\begin{array}{c}0\\1\end{array}\right)$; $\chi^{\prime (s)}=\chi^{(-s)}$. The normalization factor is given by
${\cal N}=\sqrt{\frac{E+m}{2m}}$.
Choosing $C=i\gamma_2\gamma_0$ and the usual definition of $\gamma_\mu$ with $\gamma_0= diag({\bf 1},{\bf -1})$,  where ${\bf 1}$ is a $2\times 2$ unit matrix matrix, it is easy to see that, the usual Lorentz invariant mass term $m\bar{n}n$ gives same mass for all states: $|n,+\rangle, |\bar{n},+\rangle; |n,-\rangle,|\bar{n},-\rangle$.
The $B$-violating $n-\bar{n}$ oscillation term $\delta_{B=2} n^TCn$ connects $|n,+\rangle$ with $|\bar{n},+\rangle$ and $|n,-\rangle$ with $|\bar{n},-\rangle$ in the $4 \times 4$ mass matrix written in the basis
\{$|n,+\rangle, |\bar{n},+\rangle; |n,-\rangle,|\bar{n},-\rangle$\}.

Now let us choose the $B$-conserving but Lorentz violating  mass term ${\cal L}_{LV}=\delta^{(1)}_{LV} n^\dagger n-i\delta^{(2)}_{LV} n^\dagger \partial_0n$.  Pauli matrix algebra then leads to the $4 \times 4$ mass matrix of the form\footnote{Some details of the algebra are as follows: first note that the interactions involve normal ordered product of fermionic operators;  keeping only $b^\dagger b$ and $d^\dagger d$ terms in $n^\dagger n$, we get (apart from multiplying factors): $b^\dagger b u^\dagger u - d^\dagger d v^\dagger v \sim \frac{2E}{E+m}b^\dagger b-\frac{2E}{E+m}d^\dagger d$, which on taking non-relativistic limit leads to the $\delta^{(1)}_{LV}$ term with opposite signs for $n$ and $\bar{n}$. In the calculation for the $\delta^{(2)}_{LV}$ term, the $\partial_0$ term puts in an extra negative sign so that it contributes with same sign to both $n$ and $\bar{n}$.} (in the basis \{$|n,+\rangle, |\bar{n},+\rangle; |n,->,|\bar{n},-\rangle$)\}
\begin{eqnarray}
M_{4\times 4}~=~\left(\begin{array}{cccc} m+\delta^{(1)}_{LV}+m\delta^{(2)}_{LV} & \delta_{B=2} & 0 & 0 \\ \delta_{B=2} & m-\delta^{(1)}_{LV}+m\delta^{(2)}_{LV}  & 0 & 0\\ 0 & 0 & m+\delta^{(1)}_{LV}+m\delta^{(2)}_{LV}  & -\delta_{B=2}\\ 0 & 0 & -\delta_{B=2} & m-\delta^{(1)}_{LV}+m\delta^{(2)}_{LV} \end{array}\right).\label{mat}
\end{eqnarray}
Note first that the CPT violating LIV term changes sign between the neutron and anti-neutron whereas the CPT conserving LIV term $\delta^{(2)}_{LV}$ has the same sign between the neutron and anti-neutron.
The first term will therefore split the states and we will get for the Probability of transition from one state to another as:
\begin{eqnarray}
 P_{n\to \bar{n}}= \left[\frac{ \delta^2_{B=2}}{\delta^{(1)^2}_{LV}+ \delta^2_{B=2}}\right]{\rm sin}^2\left( \frac{\sqrt{(\delta^{(1)^2}_{LV}+\delta^2_{B=2})}\,t}{\hbar}\right)e^{-\lambda t}
\end{eqnarray}
where $\lambda^{-1}  = \tau_n = 0.88 \times 10^3$ sec.

 
 If $\delta_{LV}^{(1)} \ll \delta_{B=2}$, then by assumption the current $n-\bar{n}$
oscillation limit would imply $\delta_{LV}^{(1)} \leq 10^{-28}$ GeV.  If $ \hbar/t \gg
\delta_{LV}^{(1)}\geq \delta_{B=2}$, a condition necessary to observe $n-\bar{n}$ oscillations with
free neutrons, then the first inequality would imply $\delta_{LV}^{(1)} \leq 10^{-23}
GeV$, for $t \sim 1$ sec, which is typical for the experimental setup. Note that for the current set-ups, the probability $P_{n-\bar{n}}\leq 10^{-8}$. Therefore if  $\delta_{LV}^{(1)} \gg\delta_{B=2}$, then the oscillating part averages out to $1/2$ and we get $P_{n-\bar{n}}\sim \frac{1}{2} \left[\frac{ \delta^2_{B=2}}{\delta^{(1)^2}_{LV}+ \delta^2_{B=2}}\right]$. The current limit then implies that $\frac{1}{2} \frac{ \delta^2_{B=2}}{\delta^{(1)^2}_{LV}}\leq 10^{-8}$ leading to a limit $\delta^{(1)}_{LV}\leq 10^{-24}$ GeV.

Several comments are now in order: 

(i) Note that this is less than $m^2/M_{Pl}$, where $m$ is the mass of the neutron. One could imagine such effects arising from quantum gravity effects. It is interesting that the potential limit which can be derived from $n-\bar{n}$ oscillation discovery.

(ii)  We point out that the second LIV but CPT conserving term $\delta_{LV}^{(2)}$ cannot be restricted by observation of $n-\bar{n}$ oscillation since it can be absorbed by redefinition of the mass term of the neutron and the antineutron. It is interesting that terms with this structure are induced by ambient gravitational fields like that of the Earth or the Sun.

(iii)  An important point to emphasize is that if $n-\bar{n}$ oscillation (or a $\Delta B=2$ effect) is observed in experiments searching for nucleon decay in nuclei, no limit on the Lorentz violating terms can be derived since the nuclear potential difference between and antineutrons far exceeds any Lorentz violating effect. Inside nuclei, the potential seen by the neutron and antineutron are quite different, leading to vastly different entries
 in the (1,1) and (2,2) matrix elements of Eq. (\ref{mat}), which masks the effect of LIV. This can be seen by noting that the modified  $n-\bar{n}$ transition probability in the presence of a potential difference
felt by $n$ and $\bar{n}$ is obtained from Eq. (7) by replacing $\delta_{LV}^{(1)}$ by $\delta_{LV}^{(1)} + \Delta V/2$, where $\Delta V$ is the potential difference.  The nuclear
potential difference between $n$ and $\bar{n}$ is $\Delta V \sim 100$ MeV, which would overpower the effects of LIV operator $\delta_{LV}^{(1)}$.

(iv) Finally, it is important to emphasize that since we are considering Lorentz violating effects, the specific choices we have made are frame dependent~\cite{alanframe}. However, since the neutrons are slow moving in actual experiments ($\beta \sim 10^{-3}$), our conclusions are not affected. Also we have ignored operators that are suppressed in the non-relativistic limits.
 
(v)  The limit $\delta_{LV}^{(1)} \leq 10^{-23}$ GeV that can be derived if $n-\bar{n}$ oscillation is observed, is indeed probing Lorentz violation
 occurring around the Planck scale, even when LIV operators are written down at the quark level.  The LIV effects can modify the wave functions
 of the quarks, and in the six quark operator (relevant for $n-\bar{n}$ transition), there will be a term of the type $qqqqq \delta q$ with
 a single suppression by an inverse Planck mass.  Here $\delta q$ stands for the modified quark field which contains in it Lorentz violating effects.

 \section{$\Delta B=2$ LIV Operators}

 In this section we focus on LIV as well as $B$-violating mass terms and their effect on $n-\bar{n}$ oscillation. We illustrate the effect of one operator here and take up the more generalized set of operators in a subsequent section.
 \begin{eqnarray}
 {\cal O}^{B=2}~=~iC^{5}_{\mu}n^TC\gamma_5\gamma_{\mu}n+h.c.
 \end{eqnarray}
It turns out that if we consider the neutron as an elementary fermion (i.e. in the sense of an effective field theory below the QCD scale), then the only term that makes a contribution in the vanishing momentum limit is the $C^5_\mu$ term. All other terms vanish. If we set $\langle C^5_{2}\rangle=1/2\delta^\prime_{LIV}\neq 0$, we get the following operator:
 \begin{eqnarray}
 {\cal L}_{B=2, LIV}~=~\frac{1}{2}\delta^\prime_{LIV} n^T \gamma_0\gamma_5 n+h.c.
 \end{eqnarray}
 To see its effect on the $4\times 4$ $n-\bar{n}$ matrix, let us rewrite the matrix in Eq. (2) when this is the only LIV term present, we get in the basis \{$|n,+\rangle, |\bar{n},+\rangle; |n,-\rangle,|\bar{n},-\rangle$\},
 \begin{eqnarray}
M_{4\times 4}~=~\left(\begin{array}{cccc} m & \delta_{B=2} & 0 & \delta^\prime_{LIV} \\ \delta_{B=2} & m  & \delta^\prime_{LIV} & 0\\ 0 & \delta^\prime_{LIV} & m  & \delta_{B=2}\\ \delta^\prime_{LIV} & 0 & \delta_{B=2} & m \end{array}\right).
\end{eqnarray}
In the presence of a magnetic field, this matrix becomes,
 \begin{eqnarray}
M_{4\times 4}~=~\left(\begin{array}{cccc} m+\mu B & \delta_{B=2} & 0 & \delta^\prime_{LIV} \\ \delta_{B=2} & m-\mu B  & \delta^\prime_{LIV} & 0\\ 0 & \delta^\prime_{LIV} & m-\mu B  & \delta_{B=2}\\ \delta^\prime_{LIV} & 0 & \delta_{B=2} & m+\mu B \end{array}\right).
\end{eqnarray}
Looking at the $(1,1), (1,4), (4,1), (4,4) $ entries of this matrix, we conclude that, the effect of $\delta^\prime_{LIV}$  on $n-\bar{n}$ is independent of the magnetic field unlike the $\delta_{B=2}$ term, which gets suppressed as the magnetic field is increased. Thus if we set $\delta_{B=2}$ term to zero, even before degaussing, one can extract useful information about strength of the Lorentz violating $B=2$ term for the neutrons. By the same token, since the effect of this operator is independent of the magnetic field, the current limit on $\tau_{n-\bar{n}}$ as well as limits from $n-\bar{n}$ search in nucleon decay, already puts an upper limit on $\delta^\prime_{LIV} \leq 10^{-28}$ GeV. A similar effect was noted recently in~\cite{gardner}, which discusses the more detailed implications for magnetic field. We do not discuss those aspects since our goal here is to study only the constraints on the LIV operators. 

\section{Generalization to more  LIV operators} In this section, we consider a more general set of LIV operators involving neutrons. Following ~\cite{km1}, we can write the $\Delta B = 0$ LIV operators as
\begin{eqnarray}
{\cal L}_{LIV}~=~i\bar{n}\Gamma^\mu\partial_\mu n-\bar{n}Mn
\end{eqnarray}
\begin{eqnarray}\label{LIV2}
\Gamma^\mu ~&=&~e^\mu+\gamma^\mu+c^{\nu\mu}\gamma_\nu+d^{\mu\nu}\gamma_5\gamma_\nu+ f^\nu\gamma_5+\frac{1}{2}g^{\lambda\nu\mu}\sigma_{\lambda\nu}\\\nonumber
M&=&m+a_\mu\gamma^\mu+b_\mu\gamma^5\gamma^\mu+\frac{1}{2}H^{\mu\nu}\sigma_{\mu\nu}.
\end{eqnarray}
Of these terms the terms with coefficients $a, c$ were already analyzed before. The contribution of the rest of the operators to  the diagonal elements of the $4\times 4$ $n-\bar{n}$ mass matrix is given in the table I.
\begin{table}
\begin{center}
\begin{tabular}{|c||c||c||c||c|} \hline
&$n,+$ & $\bar{n},+$ & $\bar{n},-$ & $n,-$\\ \hline
$a_\mu$ &$a_0$ &$-a_0$ & $a_0$ & $-a_0$\\ \hline
$b_\mu$ &$b_3$ &$b_3$ & $-b_3$& $-b_3$ \\ \hline
$e_\mu$ & $me_0$&$-me_0$ & $+me_0$&$-me_0$\\\hline
$f_\mu$& 0 & 0 & 0 & 0\\\hline
$d_{\mu\nu}$&$md_{03}$&$-md_{03}$&$+md_{03}$&$-md_{03}$\\\hline
$g_{\mu\nu\lambda}$&$mg_{012}$&$mg_{012}$&$-md_{012}$&$-md_{012}$\\\hline
$H_{\mu\nu}$&$H_{12}$&$-H_{12}$&$+H_{12}$&$-H_{12}$\\\hline\hline
\end{tabular}
\end{center}
\caption{Dominant contributions of the various Lorentz violating terms given in Eq. (\ref{LIV2}) to the diagonal elements of the $n-\bar{n}$ mass matrix}
\end{table}
In this table, we have ignored terms which are of order $\frac{p}{m_n}$. It is clear from the Table that $a_0, e_0, d_{03}, H_{12}$ terms are strongly bounded once the
$n-\bar{n}$ oscillation is observed. The remaining terms however are not constrained.

\section{Generalization to other $\Delta B=2$ LIV operators}
In this section, we consider a more generalized set of $\Delta B=2$ LIV operators and their effect on $n-\bar{n}$ oscillation\footnote{similar lepton number violating and Lorentz invariance violating operators for the case of neutrinos were considered in ~\cite{alan5}.}.
\begin{eqnarray}
{\cal O}^{B=2}~=~in^TC\Gamma^\prime_\mu \partial^\mu n+ n^TCM^\prime n + h.c.
\label{LIVp}
\end{eqnarray}
where
\begin{eqnarray}
\Gamma^\prime_\mu ~=~ e^\prime_\mu+c^\prime_{\mu\nu}\gamma_\nu+id^{\prime,5}_{\mu\nu}\gamma_5\gamma_{\nu}+d^\prime_{\mu\nu}\Sigma_{\mu\nu}
\end{eqnarray}
and
\begin{eqnarray}
M^\prime~=~c^\prime_\mu\gamma^\mu+H^\prime_{\mu\nu}\sigma^{\mu\nu}
\end{eqnarray}
We have not included the term already discussed in sec. IV above.
Their contributions to the off diagonal terms in the $M_{4\times 4}$ $n-\bar{n}$ mass matrix is given in Table II below:
\begin{table}
\begin{center}
\begin{tabular}{|c||c||c||c||c|} \hline
&$n+\bar{n}+$ & $n+\bar{n}-$ & $n-bar{n}-$ & $n-\bar{n}+$\\ \hline
$e^\prime_\mu,d^{\prime,5}_{\mu\nu},d^\prime_{\mu\nu}$ &$0$ &$0$ & $0$ & $0$\\
$c^\prime_\mu$, $H^\prime_{\mu\nu}$&&&&\\ \hline
$c^\prime_{00}$ &$mc^\prime_{00}$ &-$mc^\prime_{00}$& $mc^\prime_{00}$& $-mc^\prime_{00}$ \\ \hline\hline
\end{tabular}
\end{center}
\caption{Dominant contributions of the various Lorentz violating terms given in Eq. (\ref{LIVp}) to the diagonal elements of the $n-\bar{n}$ mass matrix}
\end{table}

\section{Comments}

Before concluding, we make a few observations. 

\noindent(a) Even though we have written the LIV operators in terms of the effective point like neutron approximation, we could imagine these as arising from LIV operators involving quarks. For example, written in terms of quarks fields, the operator $n^\dagger n$ could come from LIV operators of the form $u^\dagger u$. If we assume the latter as coming from quantum gravity effects, we would parameterize the expected strength for the quark bilinear as $\frac{m^2_q}{M_{P\ell}}$. The effective coefficient in front of the neutron operator then would most likely be $\sim \frac{m^2_n}{M_{P\ell}}$.

\noindent(b) As far as the origin of such Lorentz violating terms, one could speculate that they arise from Lorentz violation from a hidden sector physics. For example in a mirror world type scenario, there can exist neutron-mirror neutron mixing~\cite{BB} and if there is Lorentz violating effects on mirror neutron, it could be transmitted to the familiar neutron sector via the $n-n^\prime$ mixing.

\noindent(c) In the previous sections, we have only displayed the dominant contributions in the non-relativistic limit since the experiment is done only using very slow neutrons (i.e. with $\frac{v}{c}\sim 10^{-6}$). However one could include in the analysis also these terms and wherever appropriate, the limits on them will be less stringent by this factor.

\section{Summary} To summarize, we have pointed out that a positive signal in the search for neutron-antineutron observation would imply stringent constraints on the strengths of several kinds of baryon number conserving Lorentz violating terms for neutrons. We also find that one class of $\Delta B=2$ Lorentz invariance violating terms has the effect that it can be bounded by the non-observation of $n-\bar{n}$ oscillation and this term behaves in such a way that it is independent of the magnetic field. Thus even in the absence of degaussing, one can get useful information on the nature of baryon number violating LIV terms.

\section*{Acknowledgement} We wish to thank Gustaaf Brooijman, Susan Gardner, Yuri Kamyshkov, Alan Kostelecky, Nick Mavromatos and Mike Snow for helpful comments on the arXiv version one of this article. After this paper was posted, we were informed by Nick Mavromatos that he considered Lorentz noninvariance and decoherence effects on neutron-anti-neutron oscillation~\cite{nick}. The work of KSB is supported in part by the US Department of Energy Grant No. de-sc0010108 and  that of RNM is supported
in part by the National Science Foundation Grant No. PHY-1315155.

\end{document}